\documentclass[prd,showpacs,preprintnumbers,amsmath,amssymb,superscriptaddress,floatfix,nofootinbib]{revtex4}
\usepackage{graphicx}
\usepackage{amsmath}
\usepackage{amsfonts}
\usepackage{amssymb}
\usepackage{color}
\usepackage{multirow}
\usepackage{footnote}
\usepackage[colorlinks, citecolor=blue,anchorcolor=red,menucolor=red,linkcolor=red,filecolor=red,runcolor=red,urlcolor=blue,frenchlinks=red]{hyperref}

\begin{document}

\title{Analysis of the $ \boldsymbol{\gamma\gamma \to   D\bar D}$ reaction and the $D\bar{D}$ bound state}

\author{En Wang}
\email{wangen@zzu.edu.cn}
\affiliation{School of Physics and Microelectronics, Zhengzhou University, Zhengzhou, Henan 450001, China}
\affiliation{Department of Physics, Guangxi Normal University, Guilin 541004, China}

\author{Hong-Shen Li}
\affiliation{School of Physics and Microelectronics, Zhengzhou University, Zhengzhou, Henan 450001, China}

\author{Wei-Hong Liang}
\email{liangwh@gxnu.edu.cn}
\affiliation{Department of Physics, Guangxi Normal University, Guilin 541004, China}
\affiliation{Guangxi Key Laboratory of Nuclear Physics and Technology, Guangxi Normal University, Guilin 541004, China}

\author{Eulogio Oset}
\email{eulogio.oset@ific.uv.es}
\affiliation{Department of Physics, Guangxi Normal University, Guilin 541004, China}
\affiliation{Departamento de F\'{i}sica Te\'{o}rica and IFIC, Centro Mixto Universidad de Valencia - CSIC,
Institutos de Investigaci\'{o}n de Paterna, Aptdo. 22085, 46071 Valencia, Spain}

\begin{abstract}
In this work, we investigate the reaction of $\gamma\gamma \to D\bar{D}$, taking into account the $S$-wave $D\bar{D}$ final state interaction. By fitting to the $D\bar{D}$ invariant mass distributions measured by the Belle and {\it BABAR} Collaborations, we obtain a good reproduction of the data by means of a $D\bar{D}$ amplitude that produces a bound $D\bar{D}$ state with isospin $I=0$ close to threshold. The error bands of the fits indicate, however, that more precise data on this reaction are needed to be more assertive about the position and width of such state. 
\end{abstract}



\maketitle

\section{Introduction}

The $\chi_{c0}(2P)$ state was introduced in the Particle Data Group (PDG)~\cite{PDG2020} ($\chi_{c0}(3860)$) with $J^{PC}=0^{++}$, $M=3862^{+26+40}_{-32-13}$~MeV, and $\Gamma=201^{+154+88}_{-67~-82}$~MeV~\footnote{It should be stressed that the hypothesis of the quantum numbers $J^{PC}=0^{++}$ is favored over the $2^{++}$ hypothesis at the level of 2.5$\sigma$~\cite{exp}.} based on the single experimental measurement  on the $e^+e^- \to  J/\psi D \bar{D}$ reaction reported by the Belle  Collaboration~\cite{exp}, by looking into the $D \bar{D}$ invariant mass distribution close to threshold. These experimental data have only four points below 3900 MeV, with very large errors~\cite{exp}, and it was shown in Ref.~\cite{Wang:2019evy} that this information was not sufficient to draw any conclusion about the existence of this state. Three important facts were stressed in Ref.~\cite{Wang:2019evy}: 1) The data divided by phase space did not show any peak that would justify a claim of a state at 3860 MeV; 2) A fit of the data with a bound state of $D \bar{D}$, which has been found in Refs.~\cite{Gamermann:2006nm,Nieves:2012tt,HidalgoDuque:2012pq}, was possible, but again the uncertainties were too large to make any conclusive claim; 3) A fit of the data close to threshold using a Breit-Wigner, as discussed in Ref.~\cite{Wang:2019evy}, should be avoided. This last point has been often recalled concerning fits to data~\cite{Hanhart:2015zyp,Hyodo:2020czb}.

The question remains whether there are other data which can provide good information on the possible $D \bar{D}$ bound state. One attempt was done in Ref.~\cite{Gamermann:2007mu} using early data of the 
$e^+e^- \to J/\psi D \bar{D}$ reaction from the Belle Collaboration~\cite{Abe:2007sya}.  Although a $D\bar{D}$ bound state was found consistent with the data, the quality of these data did not allow one to be too strong on the claim of this bound state. Several reactions have been suggested, measuring $D \bar{D}$ 
mass distributions close to threshold, which can help, with good statistics, to bring an answer to this question. In Ref.~\cite{Xiao:2012iq} three methods were devised to find an answer to this problem. The 
first one is the radiative decay of the $\psi(3770)$, $\psi(3770) \to \gamma X(3700) \to \gamma \eta \eta$. 
The second one proposes the analogous reaction $\psi(4040) \to  \gamma X(3700) \to \gamma \eta \eta$, and 
the third reaction is the $e^+e^-  \to  J/\psi X(3700) \to  J/\psi \eta \eta$.
  In Ref.~\cite{Dai:2015bcc} the $B^0$ decay to the $D^0 \bar{D}^0 K^0$ reaction was suggested. The $B^+\to D^0 \bar{D}^0 K^+$ reaction has been measured by the {\it BABAR} Collaboration~\cite{Lees:2014abp} and is well reproduced in Ref.~\cite{Dai:2015bcc}, but the unmeasured $B^0 \to D^0 \bar{D}^0 K^0$ reaction was found to be more useful because it does not have the tree level contribution for $D^0 \bar{D}^0$ production and is proportional to the $D^+ D^- \to D^0\bar{D}^0$ transition amplitude which contains the bound state. In Ref.~\cite{Dai:2020yfu} the $\psi (3770) \rightarrow \gamma D^0 {\bar{D}}^0$ decay was retaken, separating the $D^+ D^-$ production from the $D^0 \bar{D}^0$ one and showing that the latter has a much bigger potential to provide valuable information concerning the existence of the $D \bar{D}$ bound state. 
The idea of the $D\bar{D}$ bound state has received a recent boost with the results of the lattice QCD calculation of Ref.~\cite{Prelovsek:2020eiw} which finds a $D\bar{D}$ bound state with  binding energy $B=4.0^{+5.0}_{-3.7}$~MeV.

 Awaiting future results from some of the suggested reactions, there are interesting data that we wish to investigate here concerning that point, and these are the $\gamma \gamma \to D \bar D $ data measured by the Belle~\cite{Uehara:2005qd} and {\it BABAR} Collaborations~\cite{Aubert:2010ab}.  In Ref.~\cite{Uehara:2005qd}, the Belle Collaboration has reported the charmonium state $X(3930)$  in the reaction of $\gamma\gamma \to D\bar{D}$, with mass $3929\pm5\pm2$~MeV and width $29\pm 10\pm 2$~MeV, which are consistent with expectations for the $\chi_{c2}(2P)$ charmonium state. Later the {\it BABAR} Collaboration has also performed the $\gamma\gamma$ production of the $D\bar{D}$ system, and the $D\bar{D}$ invariant mass distribution shows clear evidence for the $X(3930)$ state, its mass and width determined to be $M=3926.7\pm 2.7 \pm 1.1$~MeV, and $\Gamma=21.3 \pm 6.8 \pm 3.6$~MeV~\cite{Aubert:2010ab}.

On the other hand the Belle and {\it BABAR} data of Refs.~\cite{Uehara:2005qd,Aubert:2010ab}  were also used in Ref.~\cite{Guo:2012tv}, making fits with Breit-Wigner structures, to  suggest that there could be an indication of a   $\chi_{c0}(2P)$ state around 3840 MeV and a width about 200 MeV, with the warning that  "{\it More refined analysis of the data with higher statistics is definitely necessary to confirm our assertion}".  An alternative point of view concerning Ref.~\cite{Guo:2012tv} would be that obtaining a state at 3837~MeV and a width $\Gamma \simeq 221$~MeV, with a method admittedly improvable, comes to reinforce the idea that some $D\bar{D}$ state around threshold seems likely.
On the other hand, the existence of $\chi_{c0}(2P)$ with such a large width is disfavored by Ref.~\cite{Gui:2018rvv}.
 Actually we will show that the data divided by phase space does not show any peak around 3840 MeV, thus weakening the guess of Ref.~\cite{Guo:2012tv}.   A different picture was suggested in Ref.~\cite{Chen:2012wy} where the peak at 3930~MeV, associated to the $\chi_{c2}(2P)$ state in Ref.~\cite{Uehara:2005qd}, is actually a combination of the $\chi_{c0}(2P)$ and $\chi_{c2}(2P)$ with masses 3920~MeV and 3942~MeV, respectively. Thus, no structure around 3840~MeV was claimed there. In the present work we provide an alternative explanation  of the combined data of Belle and {\it BABAR}~\cite{Uehara:2005qd,Aubert:2010ab} close to threshold based on the explicit consideration of the $D \bar{D}$ final state interaction, which can shed some light on the possible $D \bar{D}$ bound state.


\section{Formalism}
\label{sec:formalism}
\subsection{$D\bar{D}$ interaction in $I=0$}

In Ref.~\cite{Gamermann:2006nm}, the $S$-wave meson-meson scattering in the charm sector was studied and a prediction for a $D\bar{D}$ bound state with isospin $I=0$ was made. In Ref.~\cite{Gamermann:2007mu}, it was found that the state with  mass $M_{D\bar{D}}=3730$~MeV, and width $\Gamma_{D\bar{D}}=30$~MeV was compatible with the data of the process $e^+e^-\to J/\psi D\bar{D}$ reported by the Belle Collaboration~\cite{Abe:2007sya}. In Ref.~\cite{Xiao:2012iq}, where three methods to detect this state were suggested, a state with $M_{D\bar{D}}=3720$~MeV and $\Gamma_{D\bar{D}}=36$~MeV was found including decays to all possible pairs of light pseudoscalars.

In this paper, we use only one channel, apart from the three channels $D^+D^-$, $D^0\bar{D}^0$, and $D_s\bar{D}_s$, which is $\eta\eta$ to account for the width of the $D\bar{D}$ bound state, as used in Refs.~\cite{Xiao:2012iq,Dai:2015bcc,Wang:2019evy}. The transition potentials  $V_{i,j}$ ($i,j=D^+D^-$, $D^0\bar{D}^0$, and $D_s\bar{D}_s$) are tabulated in Table~9 of Appendix A of Ref.~\cite{Gamermann:2006nm}, and we introduce the potentials of $\eta\eta \to D^+D^-$ and $\eta\eta \to D^0\bar D^0$ with a dimensionless strength $a_{\eta\eta}$ to give the width of the $D\bar{D}$ bound state. The transition potentials of $\eta\eta$ to $\eta\eta$ and $D_s\bar{D}_s$ are not relevant and are taken as zero. As done in Ref.~\cite{Wang:2019evy}, we will multiply the potentials $V_{D^+D^-,D_s\bar{D}_s}$ and  $V_{D^0\bar{D}^0,D_s\bar{D}_s}$ by a factor $f_{D_s\bar{D}_s}$  to stress more the cusp effect.

Then the  amplitude $t_{i,j}$ for the $i$ channel to $j$ channel can be obtained from the Bethe-Salpeter equation,
\begin{equation}
T=[1-VG]^{-1}V, \label{eq:BS}
\end{equation}
where the matrix $G$ is  diagonal with each of its elements given by  the loop function for the two particles, and we take the expression of the dimensional regularization as shown in Eq.~(31) of Ref.~\cite{Gamermann:2006nm}, where the $\mu=1500$~MeV, and the subtraction constant $\alpha$ will be taken as a free parameter.

The matrix elements $V_{ij}$ for $D\bar{D}\to D\bar{D}$, $D_s\bar{D}_s$ are basically proportional to the energy of the $D$ meson. They are based on the exchange of light vector mesons in an extension of the local hidden gauge approach~\cite{Bando:1987br,Harada:2003jx,Meissner:1987ge,Nagahiro:2008cv}, and the propagator $(q^2-m^2_V)^{-1}$ of the exchange vector is replaced by $(-m^2_V)^{-1}$ ($m_V \simeq 800$~MeV). The energy dependence of $V_{ij}$ is smooth, but the potential is attractive and produces a pole of $(1-VG)^{-1}$.

\subsection{Model for the $\gamma\gamma \to D\bar{D}$ reaction}
\begin{figure}[tbhp]
\begin{center}
\includegraphics[scale=0.6]{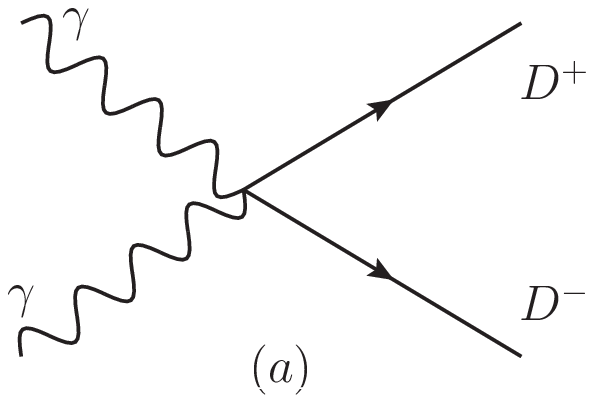}
\includegraphics[scale=0.6]{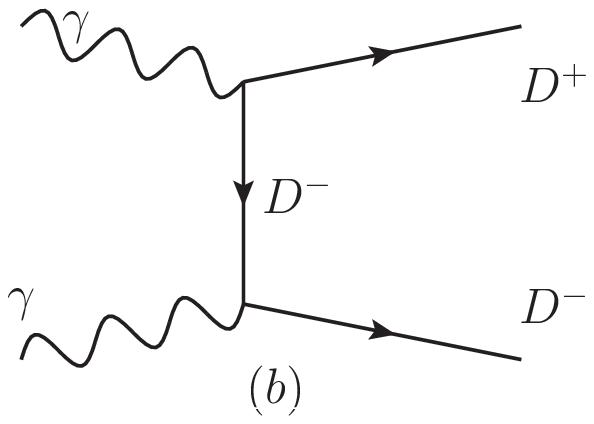}
\includegraphics[scale=0.6]{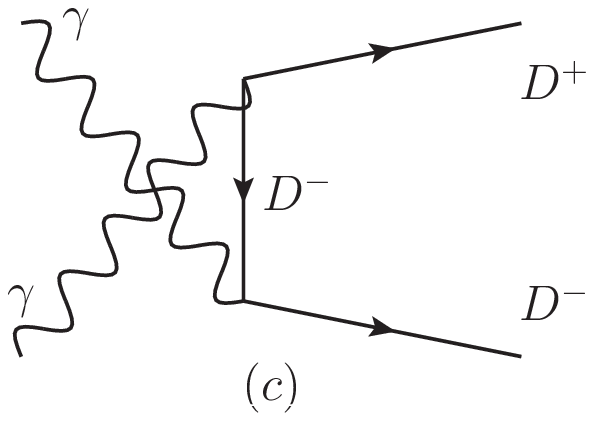}
\end{center}
\caption{Born terms for the $\gamma\gamma \to D \bar{D}$  reaction. (a) the contact term, (b) $D$ meson exchange in the $t$-channel, and (c) $D$ meson exchange in the $u$ channel.}
\label{Fig:bornterm}
\end{figure}

In this section, we  present the model for the reaction,
\begin{eqnarray}
\gamma(p,\epsilon_1) + \gamma (k,\epsilon_2)\to D^+(p')+{D}^-(k') 
\end{eqnarray} 
where $p$, $k$, $p'$, and $k'$  are  the four-momenta of the two incoming photons, $D^+$, and $D^-$, respectively, and $\epsilon_{1,2}$ are the polarizations of the two incoming photons.
We can get the mechanism for this process inspired by the work of Ref.~\cite{Oller:1997yg}, where the reactions $\gamma\gamma\to \pi^+\pi^-, \pi^0\pi^0$ were studied. In Ref.~\cite{Oller:1997yg}, the whole range of $\pi\pi$ invariant mass from 280~MeV ($2m_\pi$) till 1400~MeV was studied. The model used was good up to about 1000~MeV with no free parameters, and for higher energies the $f_2(1270)$ excitation was introduced by hand. In the present case, we only need the model for about a range of 144~MeV, from the $D\bar{D}$ threshold to 3880~MeV. The model for the process $\gamma\gamma\to D\bar{D}$ combines the Born terms: the contact term, and the $D$ meson exchange in the $t$ and $u$ channels, as shown in Fig.~\ref{Fig:bornterm}.

Contrary to the $\gamma\gamma \to \pi^+\pi^-$, the $D$-exchange terms are now here much smaller than those of $\pi$-exchange of Ref.~\cite{Oller:1997yg}, because we have the denominator in the $D$ meson propagator at threshold,
\begin{equation}
\frac{1}{(q^0)^2-(\vec{q}\,)^2-m^2_D},
\end{equation}
where $q=(q^0,\vec{q}\,)$ is the four-momentum of the exchanged $D$ meson, and we have $q^0=p^0-p^{\prime 0}=0$ and $|\vec{q}\,|=|\vec{p}\,|=p^0 \approx m_D$ at the $D\bar{D}$ threshold. So we have,
\begin{equation}
\frac{1}{0-m^2_D-m^2_D}\approx \frac{1}{-2m^2_D},
\end{equation}
which is much smaller than ${1}/({-2m^2_\pi})$ in absolute value.

These terms have also energy dependence, because we have the vertex with the term $\epsilon \cdot ({p}'-{q})$, which in the Coulomb gauge $\epsilon^0=0$ and $\vec{\epsilon}\cdot \vec{p}=0$ for the photon, which we use to evaluate, is given by,
\begin{equation}
\vec{\epsilon} \cdot \left(\vec{q}-\vec{p}^{\,\prime}\right)=\vec{\epsilon}\cdot \left( \vec{q}+ \vec{p}^{\,\prime} -2\vec{p}^{\,\prime}\right)=-2\vec{\epsilon}\cdot \vec{p}^{\,\prime}.
\end{equation}

In the limited range of the $D\bar{D}$ invariant masses that we consider, $\vec{p}^{\,\prime}$ is small and one can easily see that the contribution of the $D$-exchange terms are smaller than $3\%$ of the contact term of  Fig.~\ref{Fig:bornterm}(a), $2 e^2 \vec{\epsilon}_1\cdot \vec{\epsilon}_2$. Hence we neglect these exchange terms and take the amplitude as,
\begin{equation}
\mathcal{M}_{\gamma\gamma\to D^+D^-}= 2 e^2 \vec{\epsilon}_1\cdot \vec{\epsilon}_2. \label{eq:amp_tree}
\end{equation}
Thus, we will neglect the contributions of Figs.~\ref{Fig:bornterm}(b) and (c) in this work.

In addition, there are also other possible exchanges of $D^*$ resonances with anomalous terms but again, the denominator of the propagators are large and the terms are small close to the threshold. 

We have the differential cross section for the reaction $\gamma \gamma \to D\bar{D}$,
\begin{eqnarray}
\frac{d\sigma}{d\Omega}& =&\frac{1}{64\pi^2}\frac{1}{s}\frac{|\vec{p}^{\,\prime}|}{|\vec{p}\,|}\bar{\sum} |\mathcal{M}|^2 \nonumber \\
& =&\frac{1}{64\pi^2}\frac{1}{s}\frac{|\vec{p}^{\,\prime}|}{|\vec{p}\,|}\bar{\sum} |2 e^2\, \vec{\epsilon}_1 \cdot \vec{\epsilon}_2|^2, 
\end{eqnarray}
where we average the polarization vectors of the transverse photons, 
\begin{eqnarray}
&&\bar{\sum} \left(\vec{\epsilon}_1 \cdot \vec{\epsilon}_2\right)^2 =\frac{1}{4}\sum_{i,j} \epsilon_{1i}\epsilon_{2i}\epsilon_{1j}\epsilon_{2j} \nonumber \\
&=&\frac{1}{4}\sum_{i,j}\left(\delta_{ij}-\frac{p_{i}p_{j}}{|\vec{p}\,|^2}\right)\left(\delta_{ij}-\frac{k_{i}k_{j}}{|\vec{k}\,|^2}\right)=\frac{1}{2} 
\end{eqnarray}
with no angular dependence. Thus, we have the cross section,
\begin{equation}
\sigma =\frac{1}{8\pi}\frac{1}{s}\frac{|\vec{p}^{\,\prime}|}{|\vec{p}\,|} e^4, \label{eq:xsection}
\end{equation}
where $s=(p+k)^2$,  $\vec{p}$ and $\vec{p}^{\,\prime}$ are the three-momenta of the incoming photon and the $D^+$ in the center-mass frame, respectively,
\begin{equation} 
|\vec{p}\,|=\frac{\sqrt{s}}{2}, ~~|\vec{p}^{\,\prime}|=\frac{\lambda^{1/2}(s,m^2_D,m^2_{\bar{D}})}{2\sqrt{s}}.
\end{equation}

In our process, we have to take into account the final state interaction of the mesons $D$ and $\bar{D}$. We will differentiate between $D^+D^-$ or $D^0\bar{D}^0$, since in the experiments there is information about both. 

\subsection{Final state interaction}
\begin{figure}[tbhp]
\begin{center}
\includegraphics[scale=0.6]{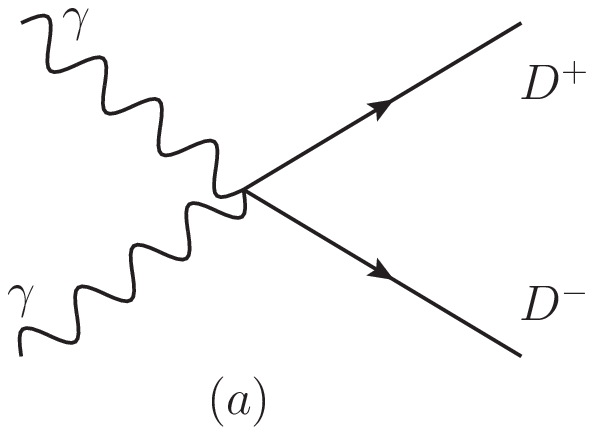}
\includegraphics[scale=0.6]{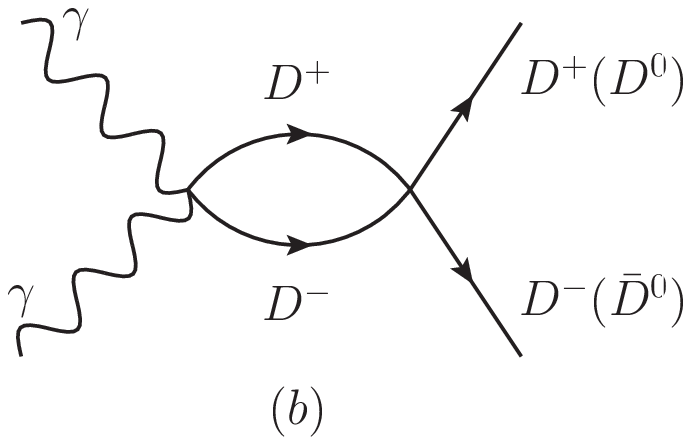}
\end{center}
\caption{The mechanism for the $\gamma\gamma \to D \bar{D}$  reaction. (a) the tree diagram, (b) the $D\bar{D}$ final state interaction.}
\label{Fig:feyn}
\end{figure}

So far we have evaluated the amplitude and cross section of the $\gamma\gamma \to D^+D^-$  at the tree level without considering the final state interaction of $D^+D^-$. We address here this problem. In addition, the $\gamma\gamma \to D^0\bar{D}^0$ is null at this level and it can only proceed via rescattering of $D^+D^-\to D^0\bar{D}^0$. This makes this reaction more favorable to learn about a possible $D\bar{D}$ bound state since the $\gamma\gamma \to D^0\bar{D}^0$ amplitude is then proportional to the $D^+D^-\to D^0\bar{D}^0$ amplitude which contains information on this possible state. The final state interaction proceeds as depicted in Fig.~\ref{Fig:feyn}(b). The amplitude in Eq.~(\ref{eq:amp_tree}) is now replaced by,
\begin{equation}
\mathcal{M}= 2 e^2 \vec{\epsilon}_1\cdot \vec{\epsilon}_2 \times t
\end{equation}
where,
\begin{eqnarray}
t_{D^+D^-}&=& 1+ G_{D\bar{D}}(M_{\rm inv})t_{D^+D^-, D^+D^-}(M_{\rm inv}), \label{eq:ampDDcharge} \\
t_{D^0\bar{D}^0}&=& G_{D\bar{D}}(M_{\rm inv})t_{D^+D^-,D^0\bar{D}^0}(M_{\rm inv}), \label{eq:ampDDneutral}
\end{eqnarray}
with $G_{D\bar{D}}$ the $D\bar{D}$ loop function and $t_{D^+D^-, D^+D^-(D^0\bar{D}^0)}$ the $D\bar{D}$ scattering amplitudes, as functions of the $D\bar{D}$ invariant mass $M_{\rm inv}$, and $t_{D^+D^-,D^+D^-}$, $t_{D^+D^-,D^0\bar{D}^0}$ the matrix elements of $T_{ij}$ in Eq.~(\ref{eq:BS}), which develop a pole when det$(1-VG)=0$. The strength of the $D\bar{D}\to D\bar{D}$ scattering matrix close to threshold is driven by the $D\bar{D}$ bound state in $I=0$~\cite{Gamermann:2006nm,Nieves:2012tt,HidalgoDuque:2012pq} and we write the $D^+D^-\to D^+D^-$ and $D^+D^-\to D^0\bar{D}^0$ scattering matrices in terms of the $D\bar{D}\to D\bar{D}$ ($I=0$) one. With the isospin doublets ($D^+$, $-D^0$), ($\bar{D}^0$, $D^-$), we have,
\begin{eqnarray}
\left| D^+D^- \right\rangle &=& \frac{1}{\sqrt{2}}\left| D\bar{D}, I=0, I_3=0\right\rangle + \frac{1}{\sqrt{2}}\left| D\bar{D}, I=1, I_3=0\right\rangle, \nonumber \\
\left| D^0\bar{D}^0 \right\rangle &=& \frac{1}{\sqrt{2}}\left| D\bar{D}, I=0, I_3=0\right\rangle - \frac{1}{\sqrt{2}}\left| D\bar{D}, I=1, I_3=0\right\rangle, \\
\end{eqnarray}
and hence,
\begin{eqnarray}
t_{D^+D^-, D^+D^-} &=& \frac{1}{2} t^{I=0}_{D\bar{D},D\bar{D}}, \nonumber \\
t_{D^+D^-, D^0\bar{D}^0} &=& \frac{1}{2} t^{I=0}_{D\bar{D},D\bar{D}}. \label{eq:transamp}
\end{eqnarray}
Equations~(\ref{eq:ampDDcharge}) and (\ref{eq:ampDDneutral}) can be rewritten as,
\begin{eqnarray}
t_{D^+D^-} &=& \left( 1+ \frac{1}{2}G_{D^+D^-}t^{I=0}_{D\bar{D},D\bar{D}} \right), \\
t_{D^0\bar{D}^0} &=&  \frac{1}{2}G_{D^+D^-}t^{I=0}_{D\bar{D},D\bar{D}} . \label{eq:amp}
\end{eqnarray}
The cross section is now given by Eq.~(\ref{eq:xsection}) multiplying it by $|t_{D^+D^-}|^2$ or $|t_{D^0\bar{D}^0}|^2$ for $D^+D^-$ or $D^0\bar{D}^0$ production, respectively.

The interpretation of the data in Refs.~\cite{Uehara:2005qd,Aubert:2010ab} requires a prior discussion. The first surprise is that in both experiments there are more events of $D^0\bar{D}^0$ production than for $D^+D^-$ production. This is surprising since the strengths of $t_{D^+D^- \to D^+D^-}$ and $t_{D^+D^- \to D^0\bar{D}^0}$ are the same (see Eq.~(\ref{eq:transamp})), but in the case of $D^+D^-$ production we have the additional tree level mechanism (see Eq.~(\ref{eq:ampDDcharge})). The answer to this question has to be seen in Table~II of Ref.~\cite{Uehara:2005qd} where the $D$ decay modes used in the detection are shown (the same detection method is used in Ref.~\cite{Aubert:2010ab}). For $D^0\bar{D}^0$ production, four decay modes are considered: 1) $D^0\to K^-\pi^+$, $\bar{D}^0\to K^+\pi^-$; 2)  $D^0\to K^-\pi^+$, $\bar{D}^0\to K^+\pi^-\pi^0$; 3)  $D^0\to K^-\pi^+$, $\bar{D}^0\to K^+\pi^-\pi^-\pi^+$; 4) $D^0\to K^-\pi^+\pi^+\pi^-$, $\bar{D}^0\to K^+\pi^-\pi^0$. However, for the $D^+D^-$ production only the $D^+\to K^-\pi^+\pi^+$, $D^-\to K^+\pi^-\pi^-$ decay mode is considered. It is thus not surprising that more $D^0\bar{D}^0$ production events than $D^+D^-$ ones are observed.  Inspection of the data in Fig.~5 of Ref.~\cite{Uehara:2005qd} shows that the strength of the $D^+D^-$ production around 3850~MeV is about $1/3$ of that of the $D^0\bar{D}^0$ production. We shall take this into account when comparing with the data. To increase the statistics, the sum of the two production modes is shown in Fig.~10 of Ref.~\cite{Aubert:2010ab}, and we shall compare with those data taking into account the experimental weights for the $D^+D^-$ and $D^0\bar{D}^0$ production. On the other hand, the data of Ref.~\cite{Uehara:2005qd} for $D^0\bar{D}^0$ production have a good statistics to compare directly with them. In view of that, in order to compare with the Belle~\cite{Uehara:2005qd} and {\it BABAR}~\cite{Aubert:2010ab} data, we shall use Eq.~(\ref{eq:xsection}) multiplied by $|t_{\rm Belle}|^2$ and $|t_{\it BABAR}|^2$, where,
\begin{eqnarray}
|t_{\rm Belle}|^2 &=& \mathcal{C}_{\rm Belle} |t_{D^0\bar{D}^0}|^2 ,\\
|t_{\it BABAR}|^2 &=& \mathcal{C}_{\it BABAR}\left( |t_{D^0\bar{D}^0}|^2 + B |t_{D^+D^-}|^2 \right) , \label{eq:ampbabar}
\end{eqnarray}
with a factor $B$ adjusted to get $\sigma_{D^+D^-}$ about $1/3$ of $\sigma_{D^0\bar{D}^0}$ around 3850~MeV. The normalization factors $\mathcal{C}_{\rm Belle}$ and $\mathcal{C}_{\it BABAR}$ are introduced to compare with the number of events in Ref.~\cite{Uehara:2005qd} and  Ref.~\cite{Aubert:2010ab}  instead of cross sections.

\section{Results and discussions}
\label{sec:res}

\begin{figure}[tbhp]
\begin{center}
\includegraphics[scale=0.6]{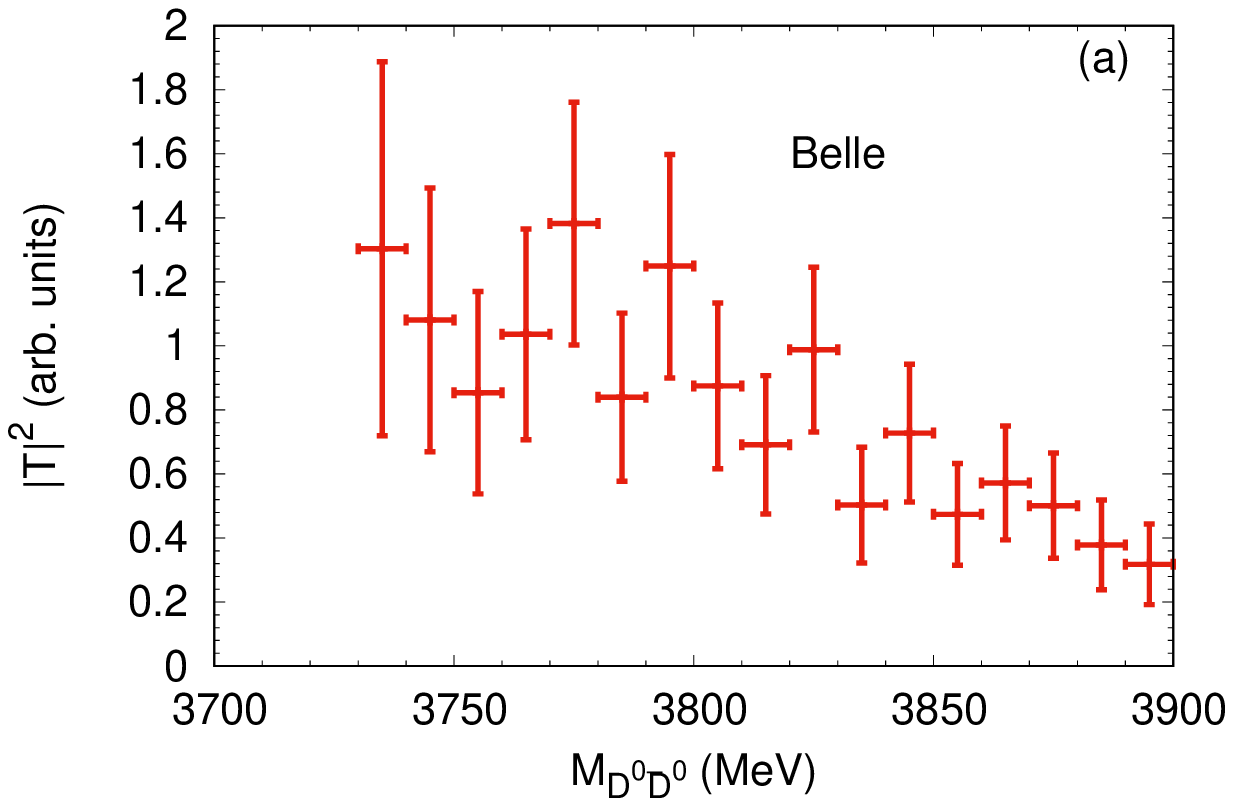}
\includegraphics[scale=0.6]{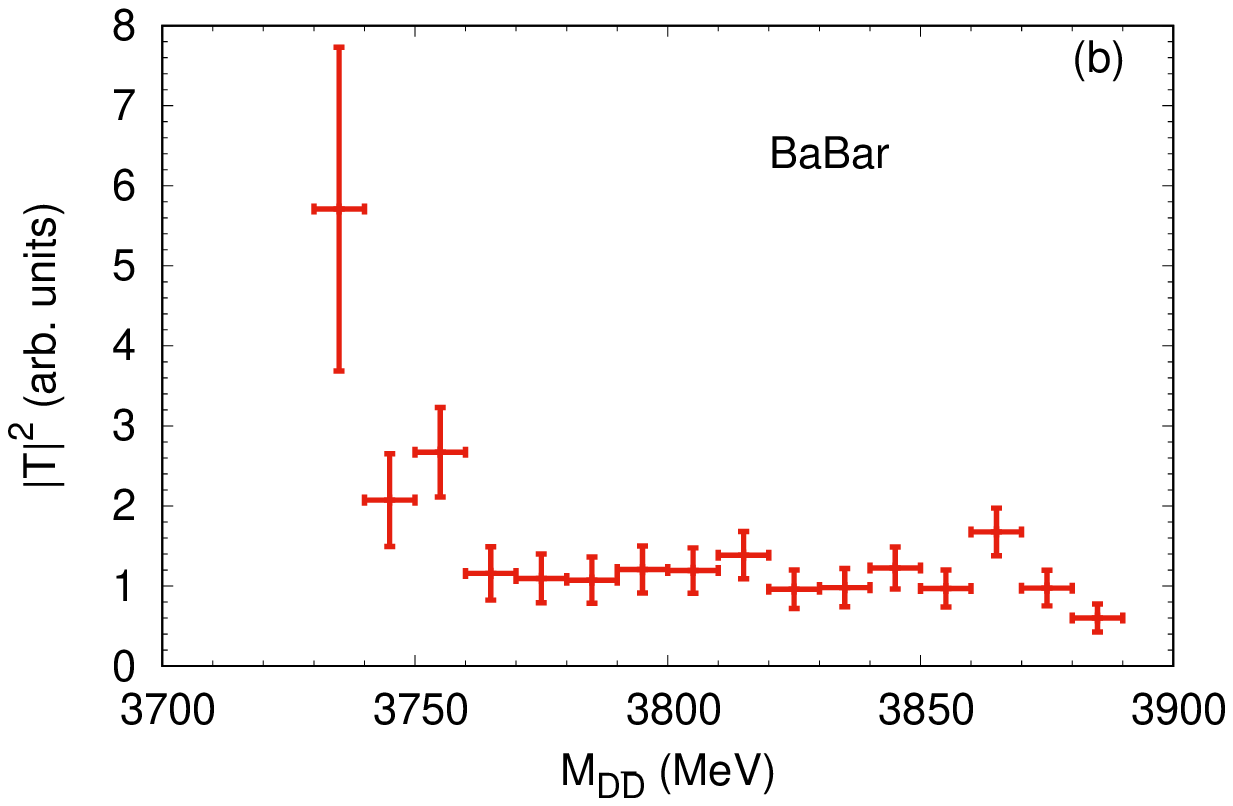}
\end{center}
\caption{The $D\bar{D}$ invariant mass distributions of $\gamma\gamma \to D \bar{D}$ measured by Belle (a) and {\it BABAR} (b) divided by the phase space  ${|\vec{p}^{\,\prime}|}/{(s|\vec{p}\,|)}$ of Eq.~(\ref{eq:xsection}). The Belle data for $\gamma\gamma \to D^0\bar{D}^0$ are taken from Fig.~2(a) of Ref.~\cite{Uehara:2005qd}, and the {\it BABAR} data for $\gamma\gamma \to D\bar{D}$ are taken from Fig.~10 of Ref.~\cite{Aubert:2010ab}. The units are in an arbitrary normalization.}
\label{Fig:reducedamp}
\end{figure}

In this section, we will show our results. 
Firstly, we divide the $D\bar{D}$ invariant mass distributions of Belle and {\it BABAR} by the phase space factor ${|\vec{p}^{\,\prime}|}/{(s|\vec{p}\,|)}$ of Eq.~(\ref{eq:xsection}), which, up to an arbitrary normalization, are shown in Fig.~\ref{Fig:reducedamp}(a) and Fig.~\ref{Fig:reducedamp}(b), respectively for the Belle and {\it BABAR} data. One can find that there are no peaks around 3860~MeV, and both distributions peak at the threshold, which implies that some possible states below the threshold may play an important role in the reaction of $\gamma\gamma\to D\bar{D}$, and a similar feature was found in the $\bar{p}\Lambda$ invariant mass distribution of $\chi_{c0}\to \bar{p}K\Lambda$~\cite{Wang:2020wap}.

\begin{table}
\begin{center}
\caption{ \label{tab:para}The model parameters obtained by fitting to the experimental measurements.}
\begin{tabular}{c  c  c  c  c  c  c c}
\hline\hline
 parameters   &$a_{\eta\eta}$       &$f_{D_s\bar{D}_s}$        &$\alpha$  & $\mathcal{C}_{\rm Belle} $  & $\mathcal{C}_{\it BABAR}$    & $\mathcal{C}$                 &$\chi^2/dof$\\
\hline
  Fit A       & $1.00\pm 0.38$   &  $2.69\pm 0.41$         & $-1.20\pm 0.04$    & $7.39\pm 0.37$     &-               & -                 &10.4/(17-4) \\
    Fit B       & $41.0 \pm 5.4$   &  $2.31\pm 0.28$         & $-1.24\pm 0.06$        &-               & $8.86 \pm 0.61$   & -               &16.3/(14-4) \\
      Fit C       & $39.1\pm 7.7$   &  $3.20\pm 0.39$         & $-1.28\pm 0.09$    & $7.85\pm 0.39$     & $8.68 \pm 0.36 $               & -                 &28.2/(31-5) \\
        Fit D       & $42.1\pm 11.0$   &  $3.30\pm 0.79$         & $-1.29\pm 0.10$     & $7.90\pm 0.41$     &$8.76\pm 0.44$               & $3.57 \pm 0.67$                 &29.9/(34-6) \\
       Fit E       & $44.1\pm 4.2$   &  $3.77\pm 0.73$         & $-1.27\pm 0.07$     & $7.90\pm 0.37$     &$8.77\pm 0.36$               & $3.58 \pm 0.68$                 &30.2/(34-6) \\

\hline\hline
\end{tabular}
\end{center}
\end{table}

\begin{figure}[tbhp]
\begin{center}
\includegraphics[scale=0.6]{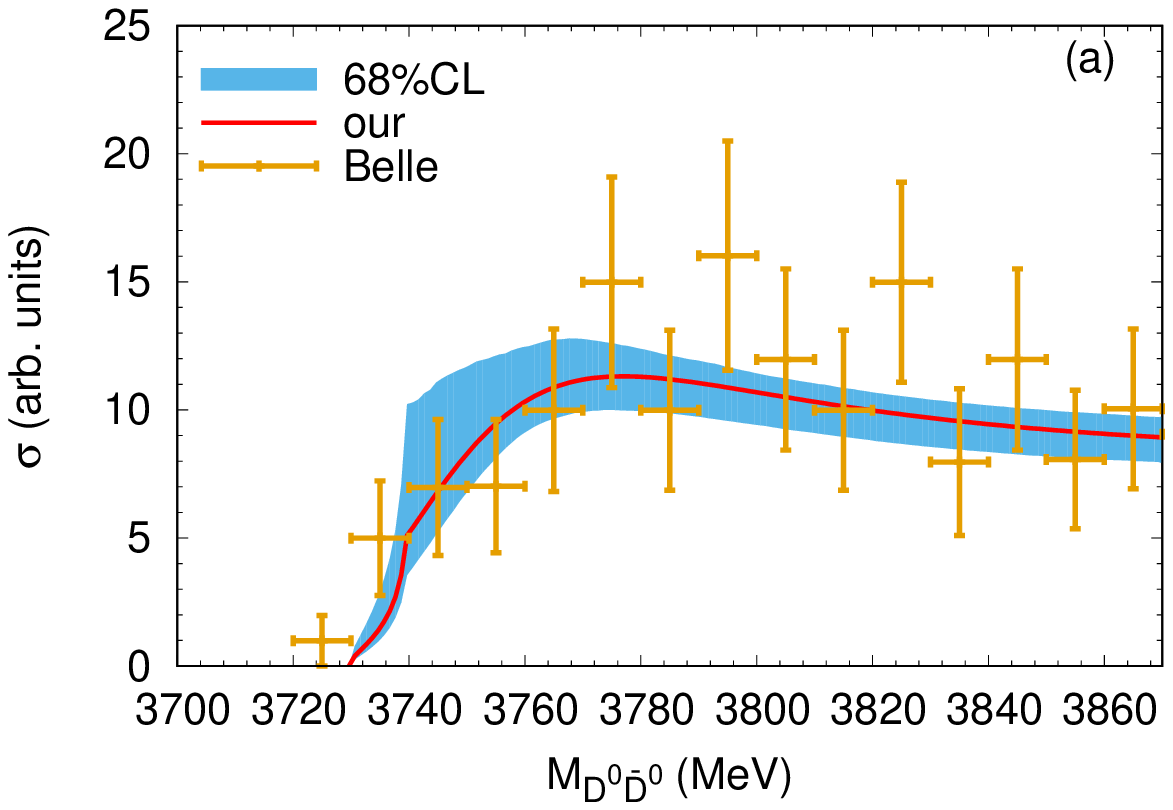}
\includegraphics[scale=0.6]{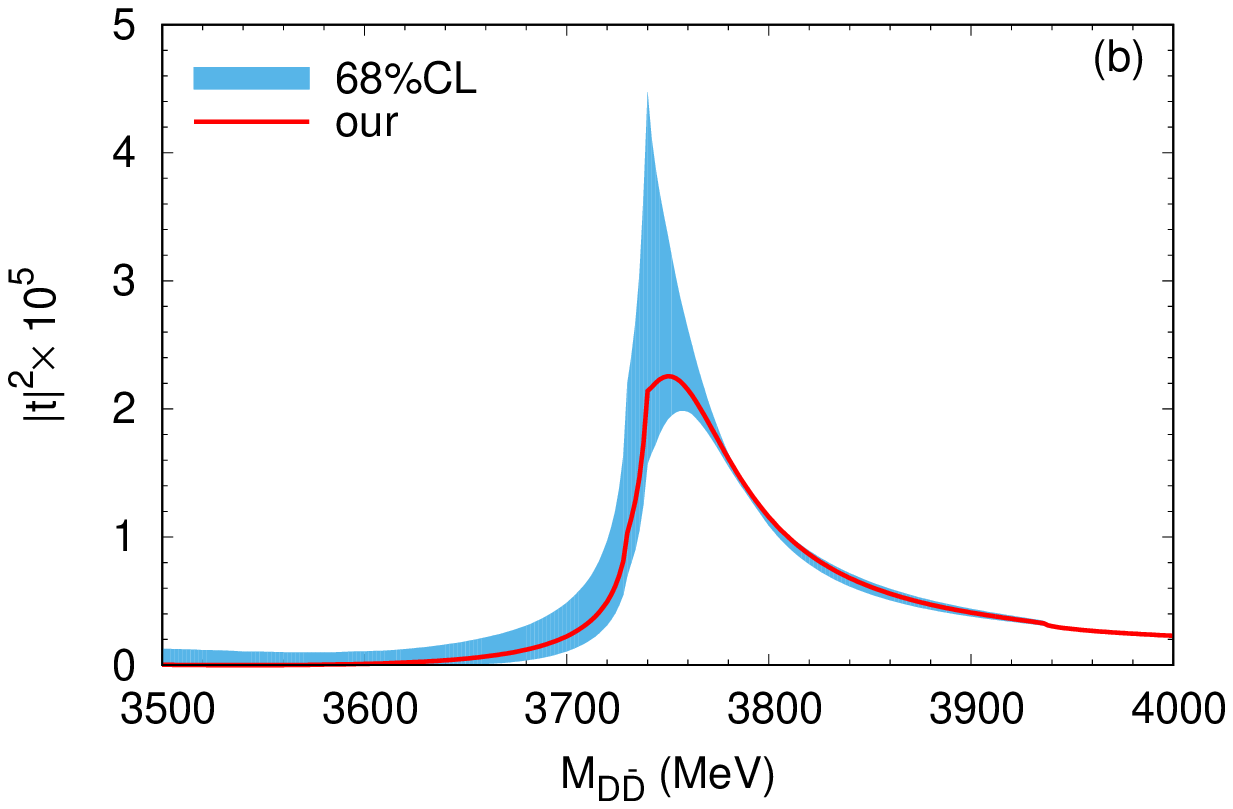}
\end{center}
\caption{(a) The mass distribution of $\gamma\gamma \to D^0 \bar{D}^0$ with the parameters fitted to the Belle data alone; (b) the modulus squared of the amplitude $|t_{D\bar{D}\to D\bar{D}}|^2$ calculated with the fitted parameters of Fit~A in Table~\ref{tab:para}. The curves labeled 'our' stand for our calculations for the $D\bar{D}$ invariant mass distribution or the modulus squared of the amplitude, the error band labeled as `68\%CL' reflects  the uncertainties from the fit and represent the 68\% confidence level, the Belle data are taken from Fig.~2(a) of Ref.~\cite{Uehara:2005qd}.}
\label{Fig:ds_belle_D0D0}
\end{figure}

\begin{figure}[tbhp]
\begin{center}
\includegraphics[scale=0.6]{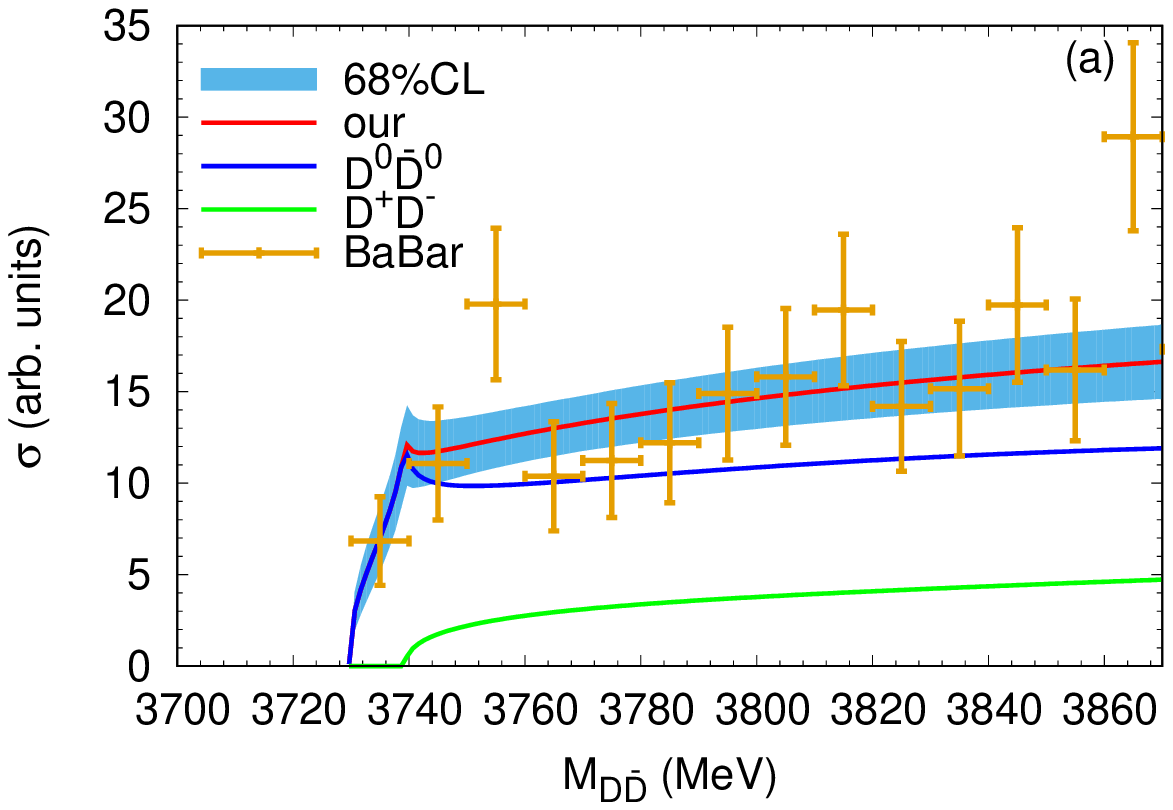}
\includegraphics[scale=0.6]{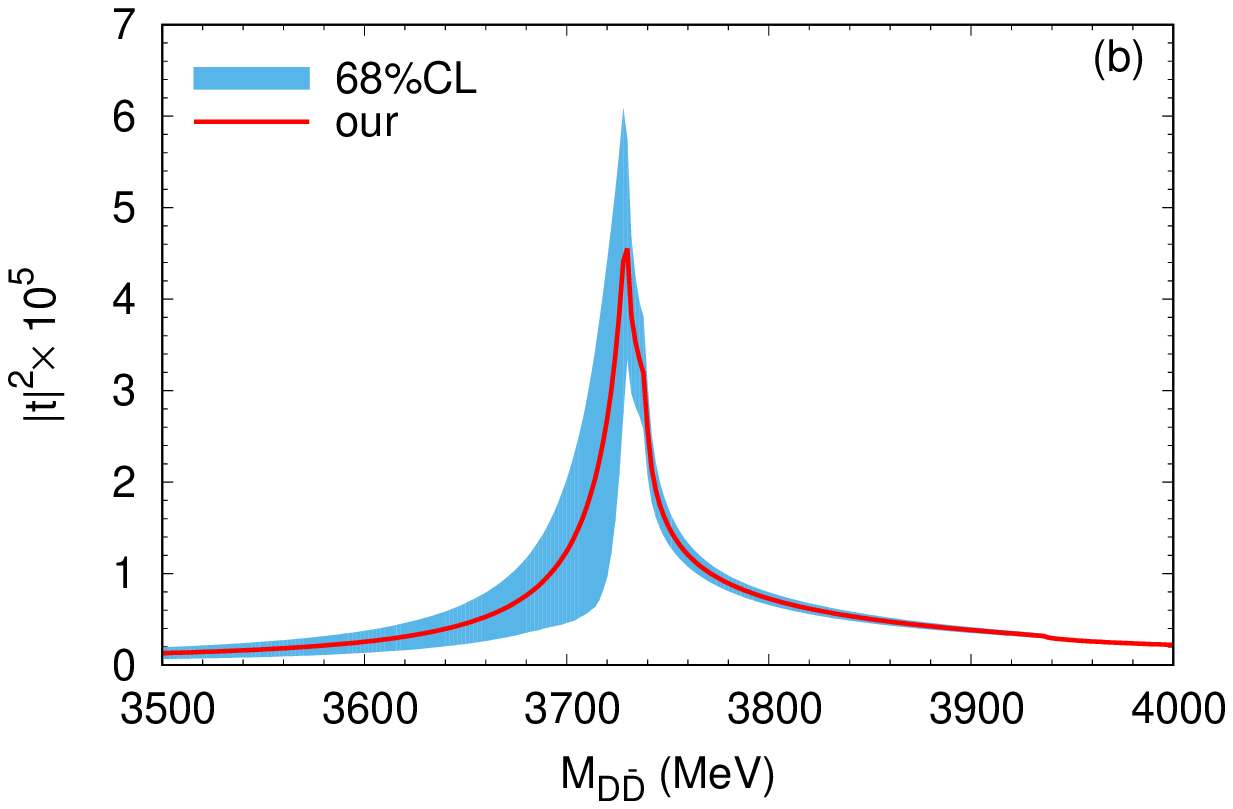}
\end{center}
\caption{(a) The mass distribution of $\gamma\gamma \to D \bar{D}$ with the parameters fitted to the {\it BABAR} data alone; (b) the modulus squared of the amplitude $|t_{D\bar{D}\to D\bar{D}}|^2$ calculated with the fitted parameters. The curves labeled as `$D^0\bar{D}^0$' and `$D^+D^-$' are the results of $\gamma\gamma\to D^0\bar{D}^0$ and $\gamma\gamma\to D^+D^-$  and the other explanation of the curves are the same as those of Fig.~\ref{Fig:ds_belle_D0D0}. The {\it BABAR} data for $\gamma\gamma \to D\bar{D}$ are taken from Fig.~10 of Ref.~\cite{Aubert:2010ab}. }
\label{Fig:ds_babar}
\end{figure}

\begin{figure}[tbhp]
\begin{center}
\includegraphics[scale=0.6]{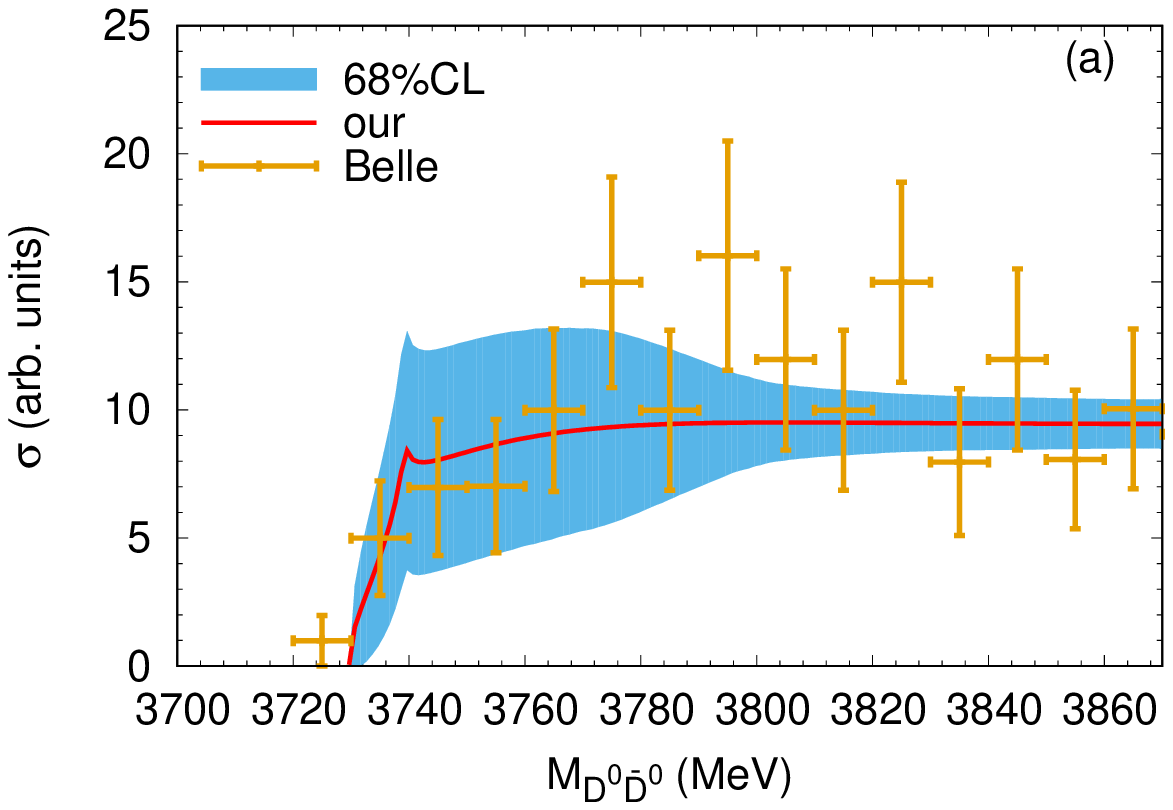}
\includegraphics[scale=0.6]{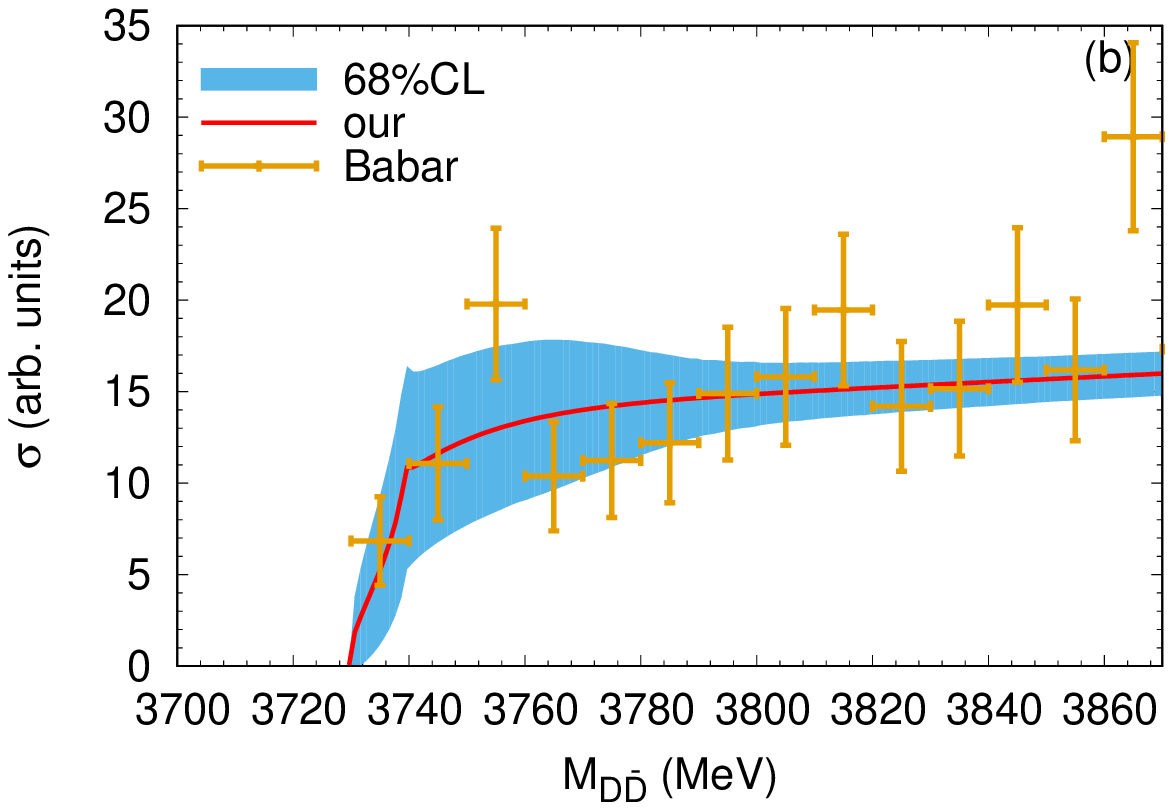}
\includegraphics[scale=0.6]{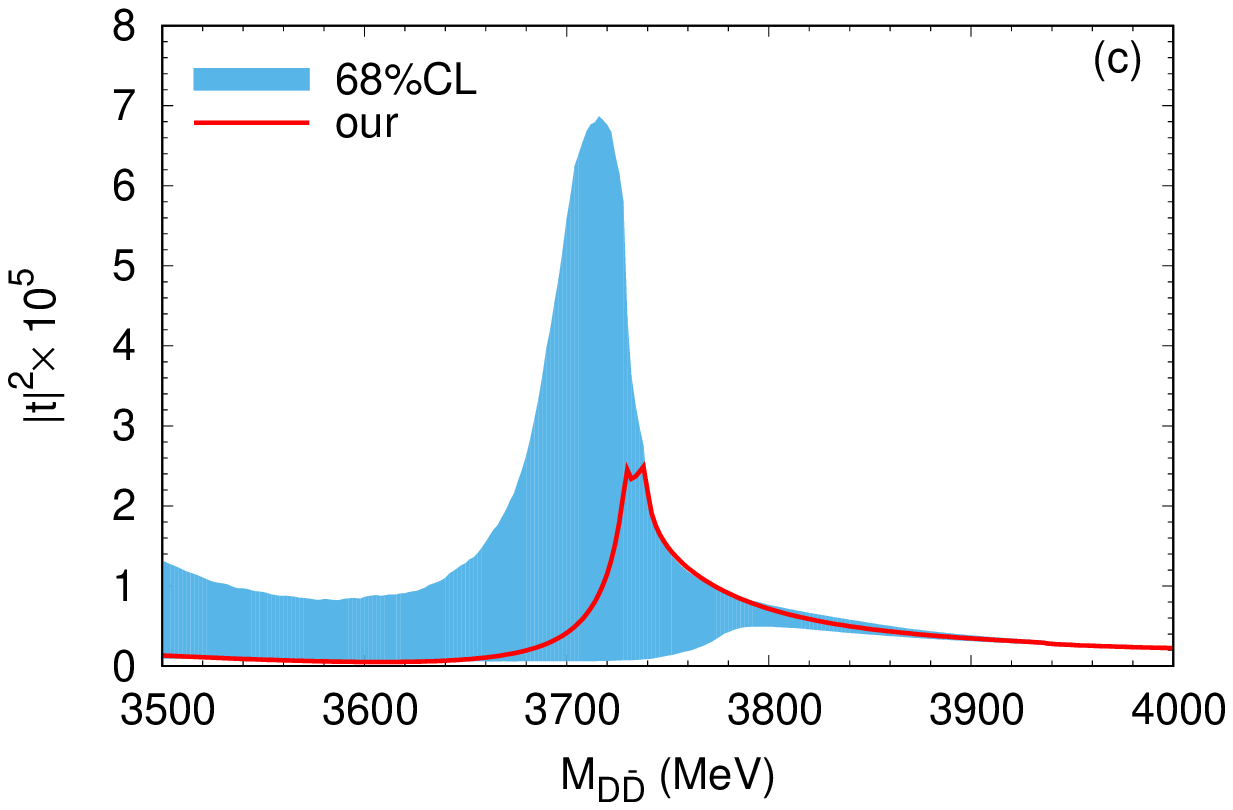}
\end{center}
\caption{The $D\bar{D}$ invariant mass distributions of $\gamma\gamma \to D \bar{D}$ with the parameters fitted to the Belle and {\it BABAR} data for (a) the $\gamma\gamma\to D^0\bar{D}^0$,  and (b) the $\gamma\gamma\to D\bar{D}$. (c) The modulus squared of the amplitude $|t_{D\bar{D}\to D\bar{D}}|^2$ calculated with the fitted parameters. The explanations of the curves are the same as those of Fig.~\ref{Fig:ds_belle_D0D0}.}
\label{Fig:ds_both}
\end{figure}

\begin{figure}[tbhp]
\begin{center}
\includegraphics[scale=0.6]{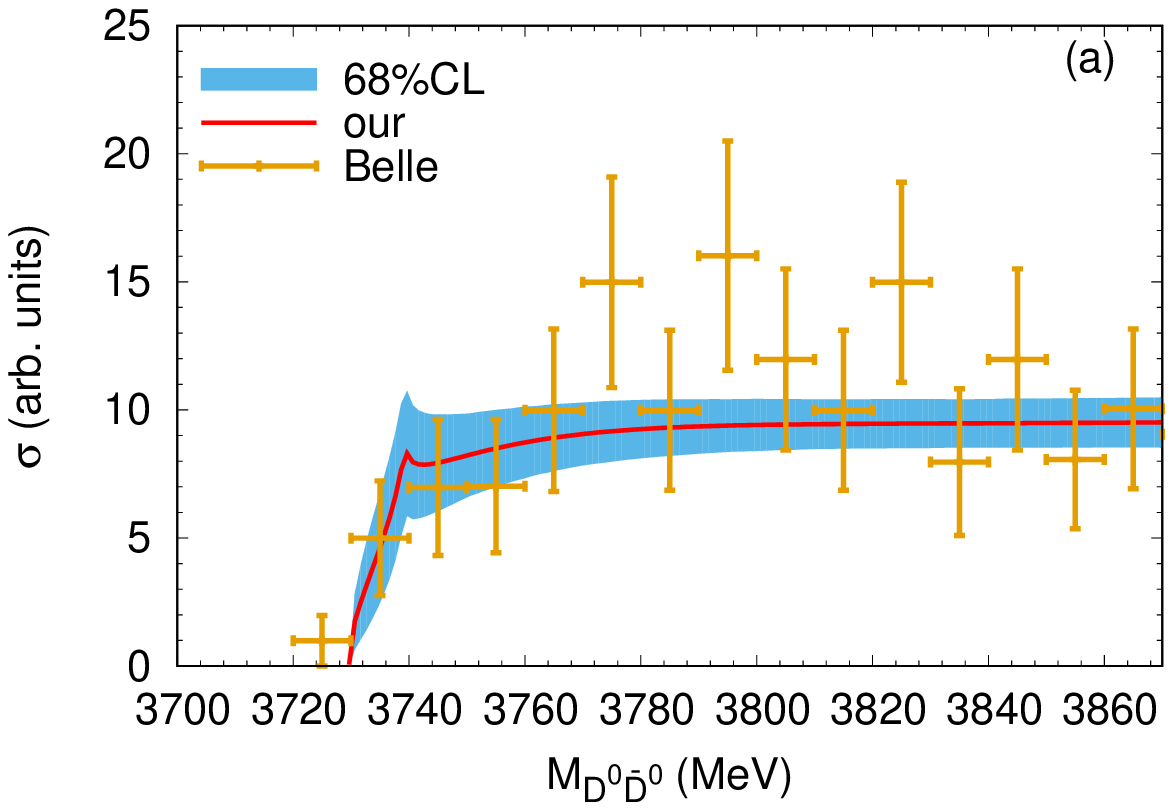}
\includegraphics[scale=0.6]{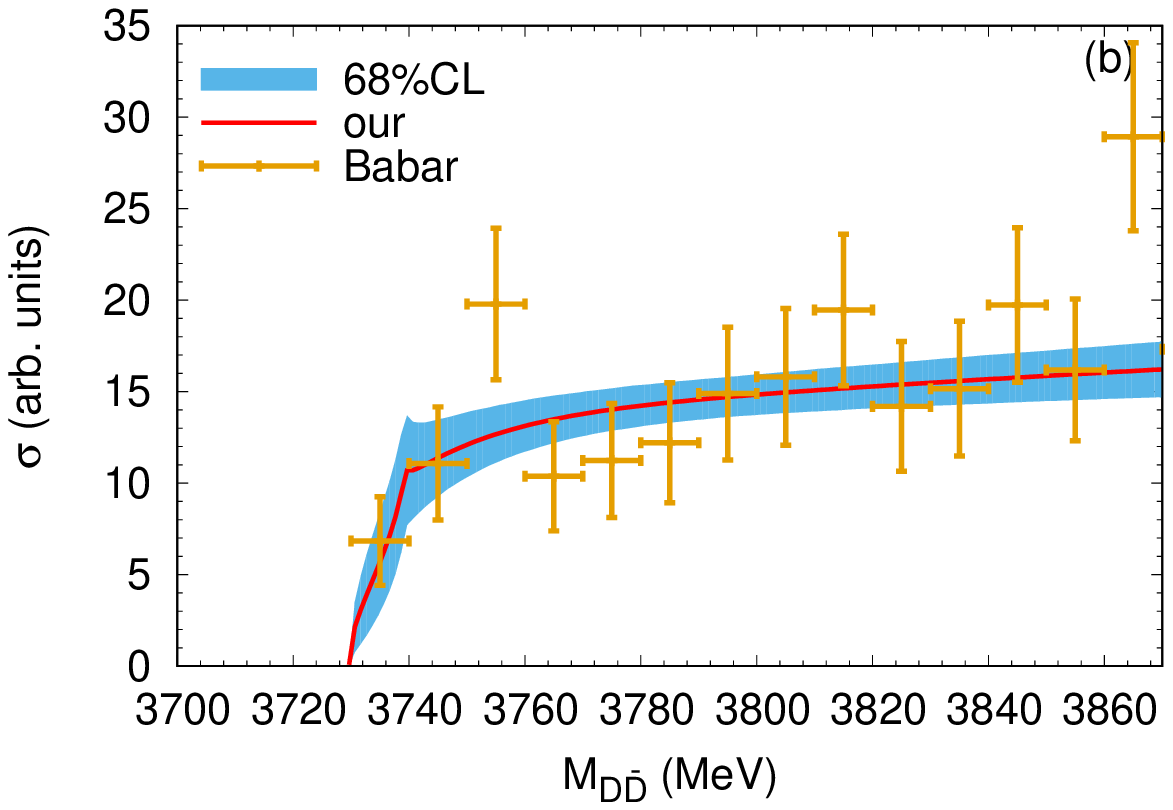}
\includegraphics[scale=0.6]{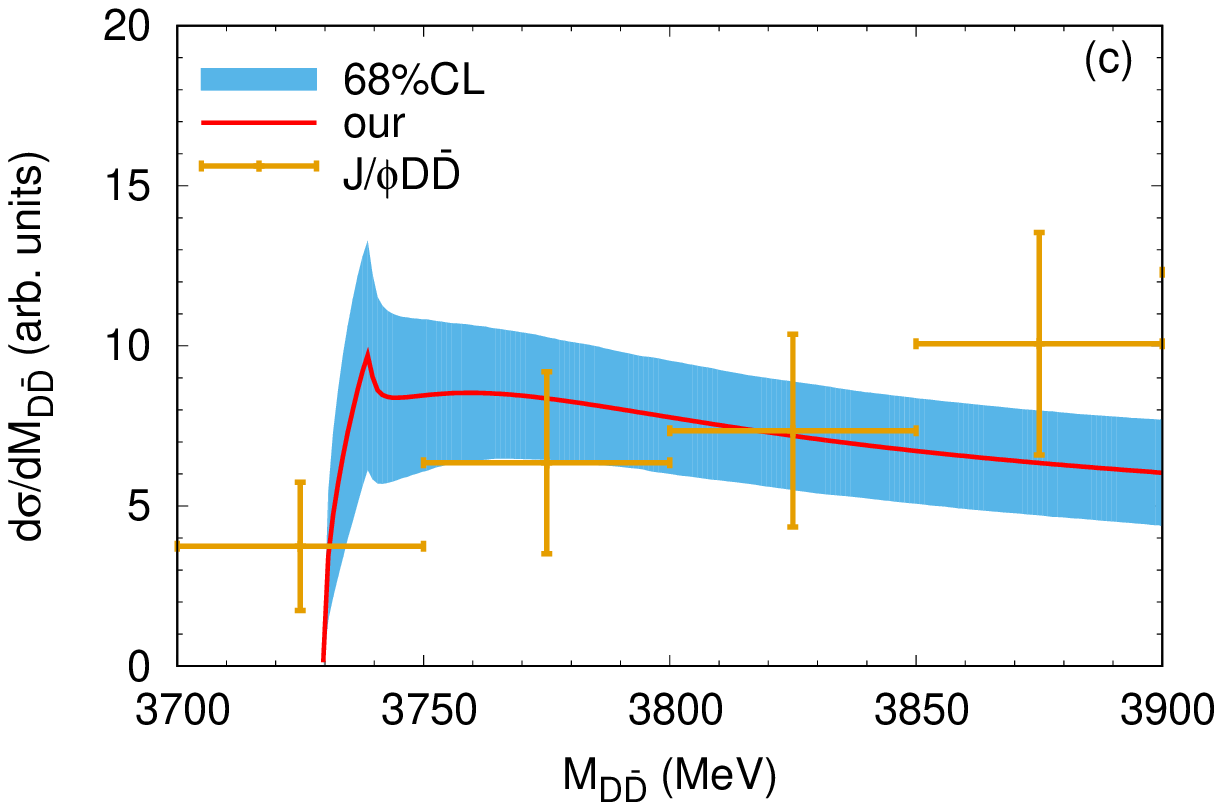}
\includegraphics[scale=0.6]{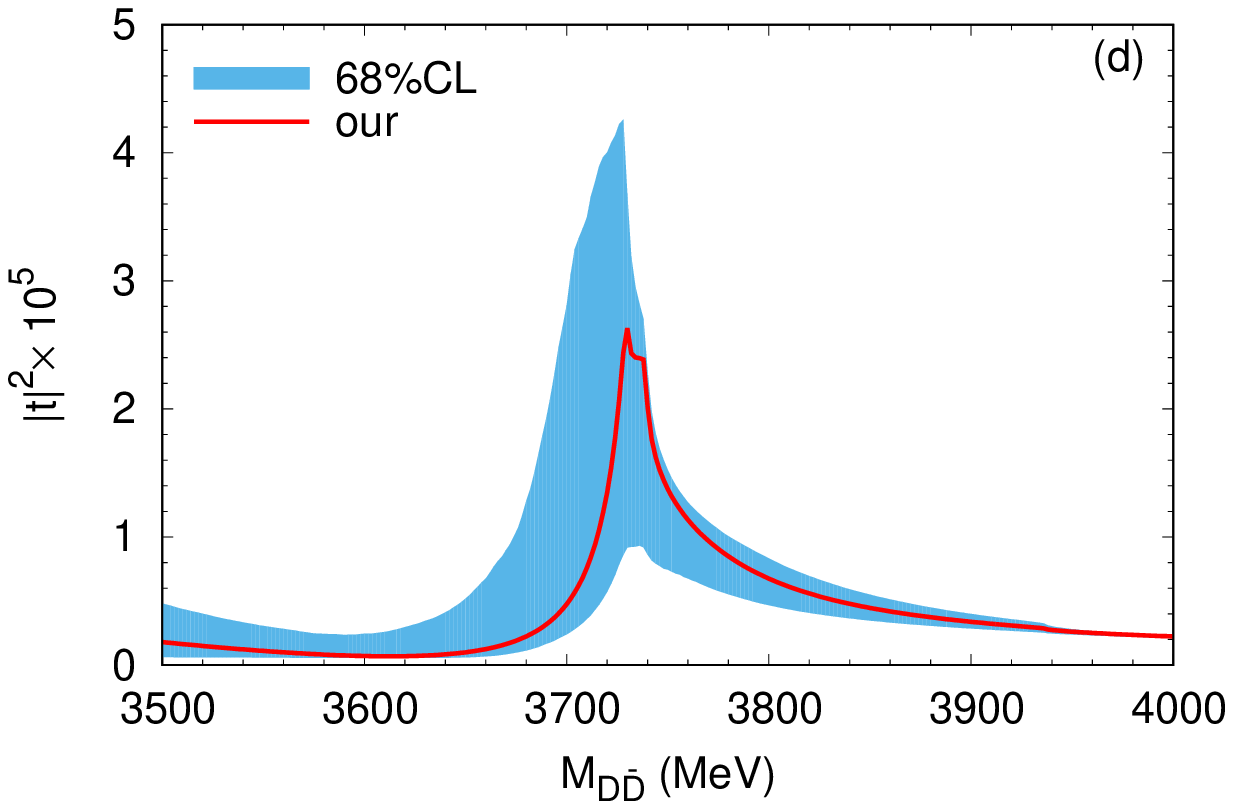}
\end{center}
\caption{The $D\bar{D}$ invariant mass distributions of $\gamma\gamma \to D \bar{D}$ with the parameters fitted to the Belle and {\it BABAR} data for (a) the $\gamma\gamma\to D^0\bar{D}^0$, (b) the $\gamma\gamma\to D\bar{D}$, and (c) $e^+e^-\to J/\psi D\bar{D}$.  (d) The modulus squared of the amplitude $|t_{D\bar{D}\to D\bar{D}}|^2$ calculated with the fitted parameters. The explanations of the curves are the same as those of Fig.~\ref{Fig:ds_belle_D0D0}. The data labeled as `$J/\psi D\bar{D}$' are taken from Ref.~\cite{exp}.}
\label{Fig:ds_three}
\end{figure}

\begin{figure}[tbhp]
\begin{center}
\includegraphics[scale=0.6]{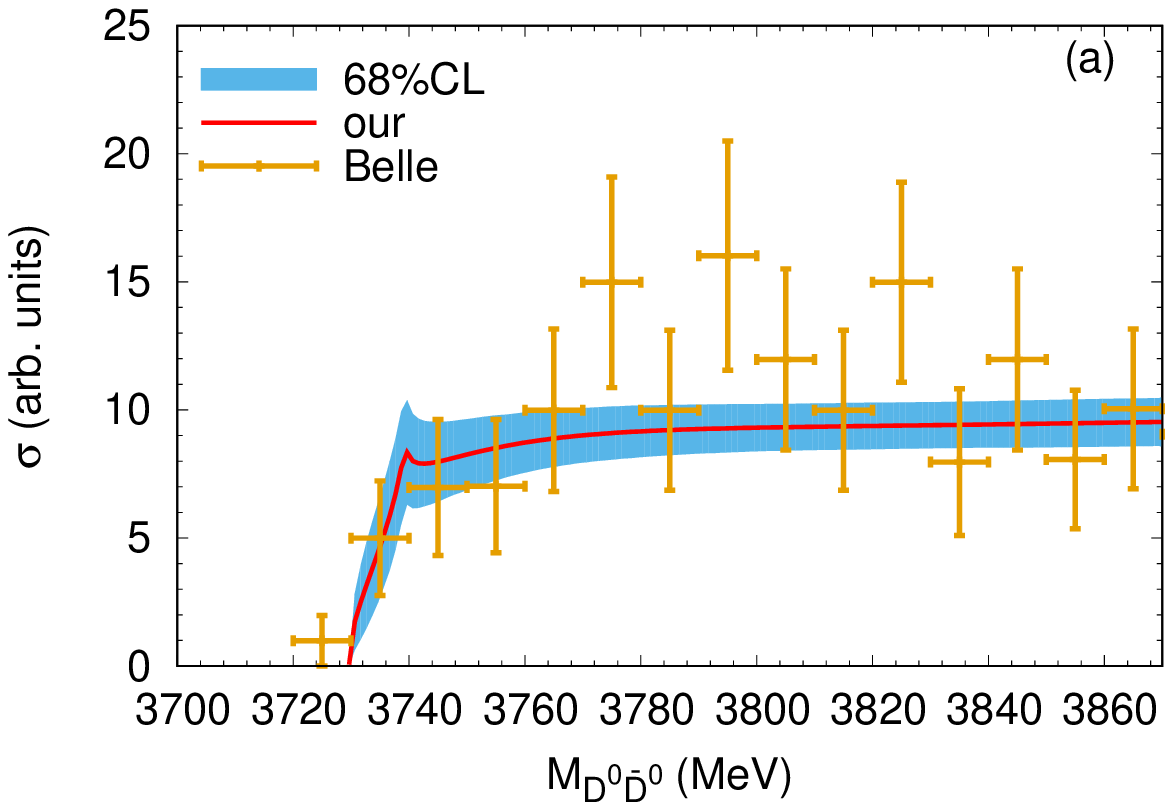}
\includegraphics[scale=0.6]{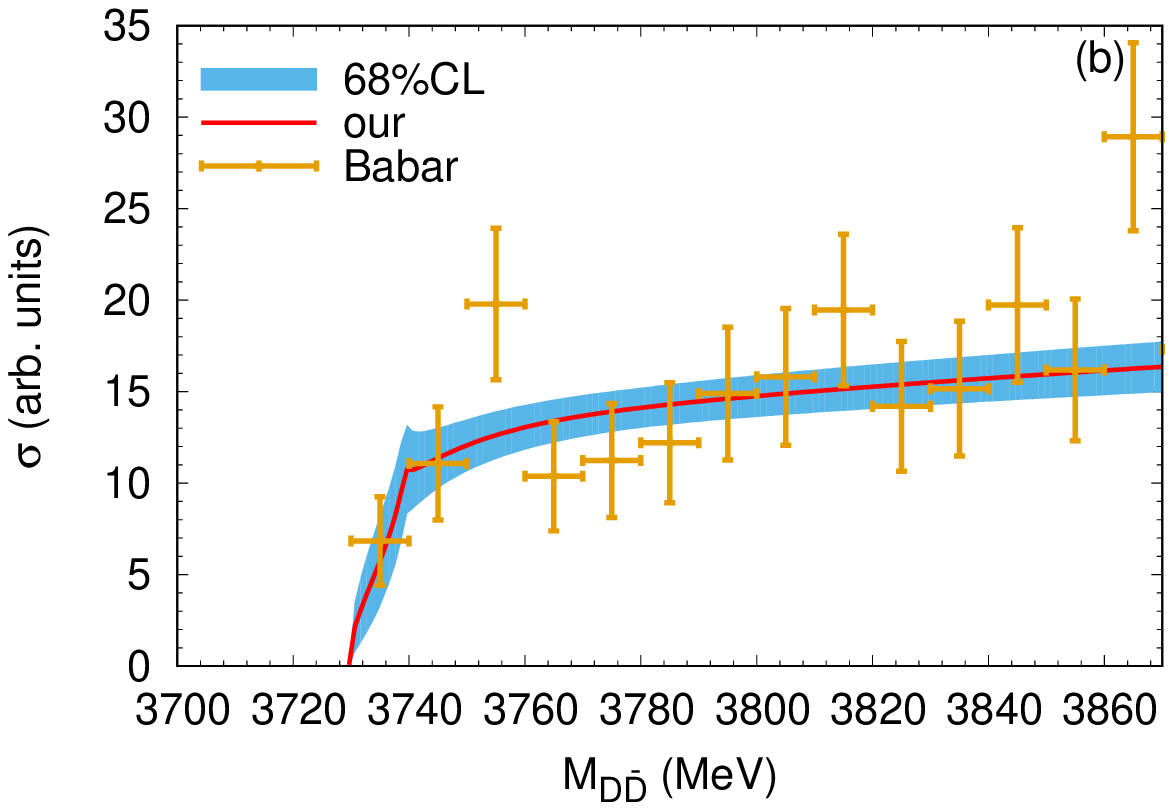}
\includegraphics[scale=0.6]{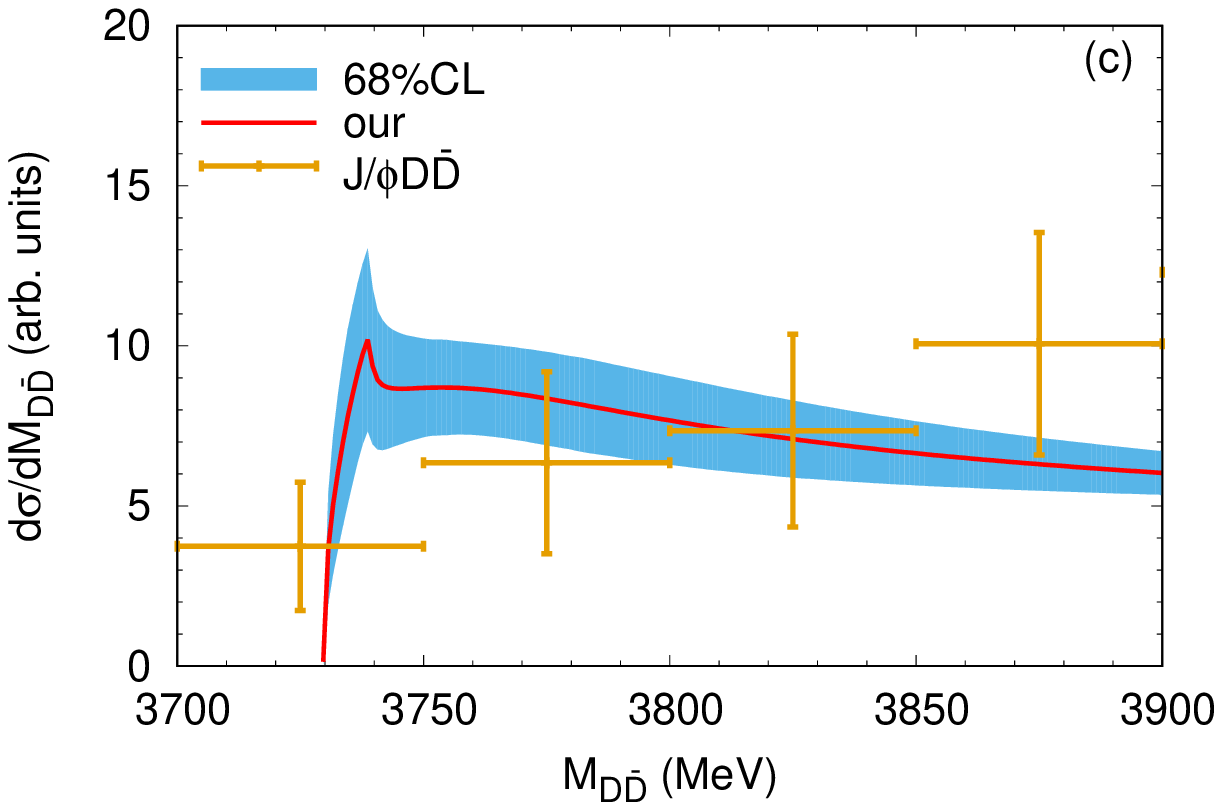}
\includegraphics[scale=0.6]{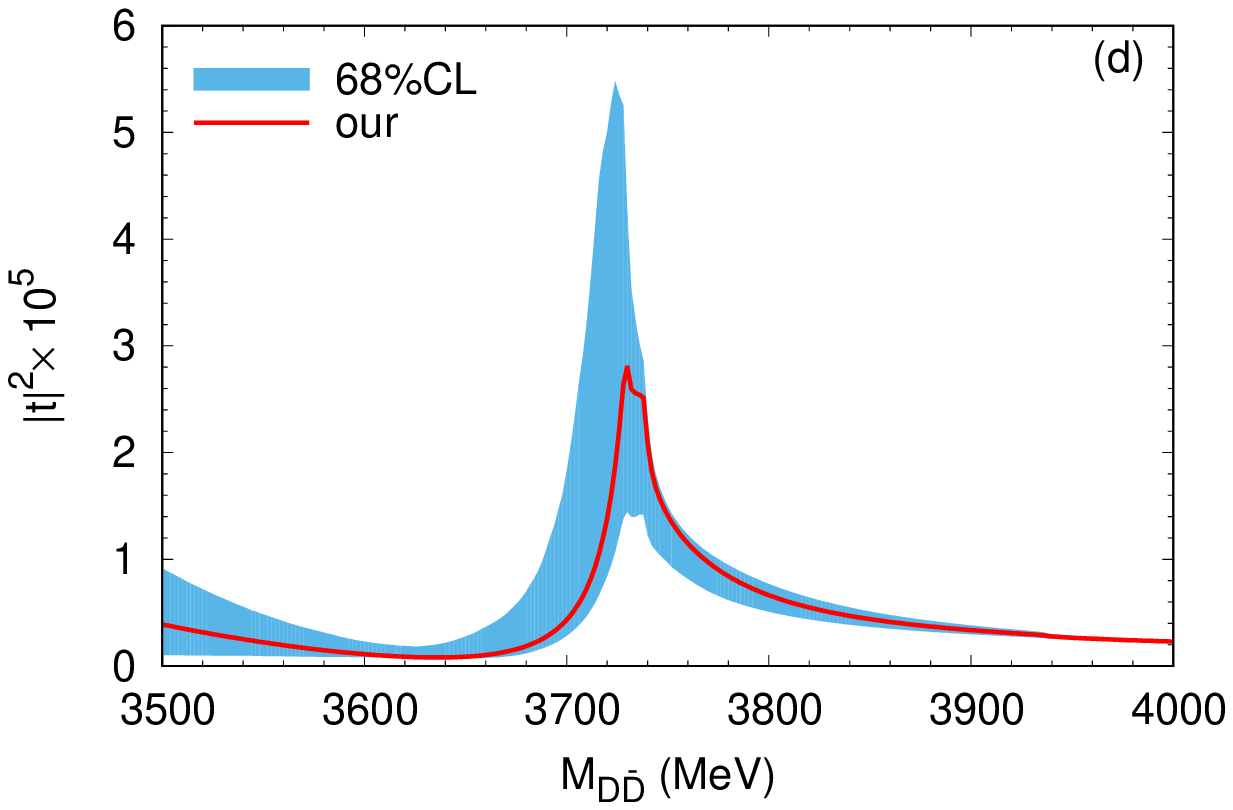}
\end{center}
\caption{The $D\bar{D}$ invariant mass distributions of $\gamma\gamma \to D \bar{D}$ with the parameters fitted to the Belle and {\it BABAR} data for (a) the $\gamma\gamma\to D^0\bar{D}^0$, (b) the $\gamma\gamma\to D\bar{D}$, and (c) $e^+e^-\to J/\psi D\bar{D}$.  (d) The modulus squared of the amplitude $|t_{D\bar{D}\to D\bar{D}}|^2$ calculated with the fitted parameters. The explanations of the curves are the same as those of Fig.~\ref{Fig:ds_belle_D0D0}. The data labeled as `$J/\psi D\bar{D}$' are taken from Ref.~\cite{exp}. In this case, we conduct an extra fit to all the data, by multiplying by 1.3 the strength of the most important potential $V_{D\bar{D},D\bar{D}}$ for the $D\bar{D}$ ($\equiv D^+D^-,D^0\bar{D}^0$) component.}
\label{Fig:ds_newfit}
\end{figure}

Let us explain this better. The experimental cross section for the reaction should be given by (see Eq.~(\ref{eq:xsection})),
\begin{equation}
\sigma =\frac{1}{8\pi}\frac{1}{s}\frac{|\vec{p}^{\,\prime}|}{|\vec{p}\,|}|t_{\gamma\gamma,D\bar{D}}|^2,
\end{equation}
where $t_{\gamma\gamma,D\bar{D}}$ is the actual $\gamma\gamma \to D\bar{D}$ transition matrix. By dividing the experimental cross section by the phase factor ${|\vec{p}^{\,\prime}|}/({s|\vec{p}\,|})$ we are isolating $|t_{\gamma\gamma,D\bar{D}}|^2$. Should this matrix contain a resonance it should show up in these data divided by the phase space.

As discussed above, there are five parameters: 1), $a_{\eta\eta}$ the dimensionless potential of $\eta\eta \to D^+D^-$ and $\eta\eta\to D^0\bar{D}^0$; 2), an extra factor $f_{D_s\bar{D}_s}$ of the potentials $V_{D^+D^-,D_s\bar{D}_s}$ and  $V_{D^0\bar{D}^0,D_s\bar{D}_s}$; 3), the subtraction constant $\alpha$ in the loop function; 4), two normalization factors $\mathcal{C}_{\rm Belle}$ and $\mathcal{C}_{\it BABAR}$. We will fit these parameters to the experimental data in the following.  It should be noted that the amplitudes produced by our model have a limited range of validity and should not be used much above the $D_s\bar{D}_s$ threshold (3937~MeV), thus we only consider the experimental data points from the $D\bar{D}$ threshold to 3860~MeV. 

In the first step, we fit to the Belle data of $\gamma \gamma \to D^0\bar{D}^0$ alone ({\bf Fit A}). The fitted parameters are tabulated in Table~\ref{tab:para}, and the mass distribution is shown in Fig.~\ref{Fig:ds_belle_D0D0}(a).  Our results are in good agreement with the Belle data of $\gamma \gamma \to D^0\bar{D}^0$. With the fitted parameters, the modulus squared of the amplitude $|t_{D\bar{D}\to D\bar{D}}|^2$ is depicted in Fig.~\ref{Fig:ds_belle_D0D0}(b), where we can find that there is a peak around $3730\sim 3740$~MeV, associated to a bound $D\bar{D}$ state.

Next we perform the fit to the {\it BABAR} data of $\gamma \gamma \to D\bar{D}$ alone ({\bf Fit B}), and the fitted parameters are tabulated in Table~\ref{tab:para}. With the fitted parameters, we show the $D\bar{D}$ mass distribution and the modulus squared of the amplitude $|t_{D\bar{D}\to D\bar{D}}|^2$ in Fig.~\ref{Fig:ds_babar}(a) and Fig.~\ref{Fig:ds_babar}(b). We have adjusted the relative weight $B$ of Eq.~(\ref{eq:ampbabar}) to get $\sigma_{D^+D^-}$ about 1/3 of $\sigma_{D^0\bar{D}^0}$ around 3850~MeV in this case and also in the following fits. It is easy to see that there is a peak around 3720 MeV, which can also be associated to the $D\bar{D}$ bound state.

Then we perform the fit to both the Belle and  {\it BABAR} data ({\bf Fit C}), and the fitted parameters are tabulated in Table~\ref{tab:para}. We present the $D\bar{D}$ invariant mass distributions in Fig.~\ref{Fig:ds_both}(a) and Fig.~\ref{Fig:ds_both}(b), respectively for the Belle and {\it BABAR} data. With the fitted parameters, the modulus squared of the amplitude $|t_{D\bar{D}\to D\bar{D}}|^2$ is given in Fig.~\ref{Fig:ds_both}(c). Taking into account the uncertainties, our results are in reasonable agreement with the Belle and {\it BABAR} measurements, and the fit favors a narrow bound $D\bar{D}$ state, which can be seen from Fig.~\ref{Fig:ds_both}(c).

As we discussed in Ref.~\cite{Wang:2019evy}, the present quality of the $e^+e^-\to J/\psi D\bar{D}$ data from the Belle Collaboration~\cite{exp} did not allow one to be too strong on the claim of a $D\bar{D}$ bound state around 3720~MeV, although this $D\bar{D}$ bound state was found to be compatible with the Belle measurements.  Since the $D\bar{D}$ final state of the $e^+e^-\to J/\psi D\bar{D}$ reaction is the same as the one of $\gamma\gamma \to D\bar{D}$, we make a global fit to the data of $\gamma\gamma\to D^0\bar{D}^0$ of Belle~\cite{Uehara:2005qd}, $\gamma\gamma\to D\bar{D}$ of {\it BABAR}~\cite{Aubert:2010ab}, and $e^+e^-\to J/\psi D\bar{D}$ of Belle~\cite{exp}~\footnote{The formalism for the 
$e^+e^-\to J/\psi D\bar{D}$ can be found in Ref.~\cite{Wang:2019evy}. In addition to the three parameters, $a_{\eta\eta}$, $\alpha$, $f_{D_s\bar{D}_s}$, we have another parameter $\mathcal{C}$ corresponding to the normalization factor in Eq.~(1) of  Ref.~\cite{Wang:2019evy}.} ({\bf Fit D}), and the fitted parameters are tabulated in Table~\ref{tab:para}. The mass distributions of $\gamma \gamma \to D\bar{D}$ are shown in Fig.~\ref{Fig:ds_three}(a) and Fig.~\ref{Fig:ds_three}(b) for Belle and {\it BABAR}, respectively. The $D\bar{D}$ mass distribution of $e^+e^-\to J/\psi D\bar{D}$ is shown in Fig.~\ref{Fig:ds_three}(c). With the fitted parameters, the modulus squared of the amplitude $|t_{D\bar{D}\to D\bar{D}}|^2$ is given in  Fig.~\ref{Fig:ds_three}(d).  The global fit also favors a $D\bar{D}$ bound state around 3720~MeV. 

In order to find uncertainties in our model, we conduct an extra fit to all the data, by multiplying by 1.3 the strength of the most important potential $V_{D\bar{D},D\bar{D}}$ for the $D\bar{D}$ ($\equiv D^+D^-,D^0\bar{D}^0$) component. The results are shown in Table~\ref{tab:para} as {\bf Fit E} and in Fig.~\ref{Fig:ds_newfit}. We see that there are only small difference to the former {\bf Fit D}, as a consequence of a well known fact that changes in the potentials can be accommodated to a large degree by some change in the subtraction constant.

\section{Conclusions and Perspective}
\label{sec:conc}
In this work, we have investigated the reaction of $\gamma\gamma \to D\bar{D}$ by taking into account the $S$-wave $D\bar{D}$ final state interactions.  Since the present quality of the $e^+e^-\to J/\psi D\bar{D}$ data from the Belle Collaboration did not allow one to be too strong on the claim of the $D\bar{D}$ bound state, and the final states of $e^+e^-\to J/\psi D\bar{D}$ and $\gamma\gamma \to D\bar{D}$ are the same, we perform five kinds of fits to the data of $\gamma \gamma \to D^0\bar{D}^0$ from the Belle Collaboration,  $\gamma\gamma \to D\bar{D}$ from the {\it BABAR} Collaboration, and $e^+e^-\to J/\psi D\bar{D}$ from the Belle Collaboration. Considering the uncertainties from the fitted parameters, our results are consistent with the experimental data in the four fits, and the modulus squared of the amplitudes $|t_{D\bar{D}\to D\bar{D}}|^2$ show peaks around $3710\sim 3740$~MeV, which can be associated to the $D\bar{D}$ bound state. Yet, the explicit evaluation of the errors done in each of the fits to the data, and particularly the last two including all the data, show that there are still large uncertainties to be assertive about the position and width of the state.

We would like to call the attention to other possible uncertainties in our theoretical approach. We have assumed that our bound state of $D\bar{D}$ is a pure bound state appearing from the meson-meson interaction and hence it qualifies as a pure molecular state. There is the issue of possible mixing with ordinary $c\bar{c}$ states. This issue is relevant and has been addressed formally in Ref.~\cite{Cincioglu:2016fkm} for the $D\bar{D}$ case, and with numerical results for the $D\bar{D}^*$ ($1^{++}$), $D^*\bar{D}^*$ ($2^{++}$) systems in Ref.~\cite{Cincioglu:2016fkm} and for the $D\bar{D}^*$ ($1^{++}$) system in Ref.~\cite{Ortega:2009hj}. In this latter work which uses quark dynamics, three states are found at 3489~MeV, 3871~MeV, and 3942~MeV. The first state has $97\%$ component of the $c\bar{c}$ ($1^3P_1$) state and $3\%$ of $D\bar{D}^*$. The  second state has $7\%$ of $c\bar{c}$ ($2^3P_1$) component and $93\%$ of $D\bar{D}^*$. The last state has $88\%$ $c\bar{c}$ ($2^3P_1$) component and $12\%$ of $D\bar{D}^*$. Similar results are obtained in Ref.~\cite{Cincioglu:2016fkm} depending somehow on an unknown mixing parameter. The conclusion is that the $X(3872)$ is largely a state of $D\bar{D}^*$ nature. 

In our case ($0^{++}$), there is a $c\bar{c}$ state of $1^3P_0$ ($\chi_{c0}(1P)$) nature at 3415~MeV in the PDG. The molecular state analogous to the $X(3872)$ is the $D\bar{D}$ bound state that we get, $X(3700)$, and the $c\bar{c}$ $2^3P_0$ ($\chi_{c0}(2P)$) state could be the one obtained in Ref.~\cite{exp} at 3862~MeV. Since we question the interpretation of Ref.~\cite{exp}, to continue the discussion we recall the predictions of the relativized quark model of Ref.~\cite{Godfrey:1985xj}, where the $1^3P_0$ ($\chi_{c0}(1P)$) is predicted at 3440~MeV, close to the experimental one of 3415~MeV, and the $2^3P_0$ ($\chi_{c0}(2P)$) is predicted at 3920~MeV. Taking the latter number as reference, there is a separation of 200~MeV between this state and our predicted $D\bar{D}$ bound state, much bigger than the 70~MeV that one has between the $X(3872)$ and the $2^3P_1$ state at 3942~MeV. Given the larger separation of the state  and the small mixture found in Ref.~\cite{Ortega:2009hj} for the $1^{++}$ states, it is logical to think that the mixing of the $D\bar{D}$ bound state that we find and the $2^3P_0$ state would be even smaller. However, this and other considerations will have to be taken into account in the future when high precision experimental data are available. 

The recent lattice QCD results are also relevant in this context. Indeed, in the study of Ref.~\cite{Prelovsek:2020eiw} $c\bar{c}$, $D\bar{D}$, $D_s\bar{D}_s$ interpolators are used and several states are obtained. The $\chi_{c0}(1P)$ and $\chi_{c2}(1P)$ state are obtained in good agreement with experiments, coupling mostly to the $c\bar{c}$ components. In addition, a second $\chi_{c2}$ state is obtained which can be associated to the $\chi_{c2}(3930)$ ($\chi_{c2}(2P)$). In the $\chi_{c0}$ $(0^{++})$ sector, in addition to the $\chi_{c0}(1P)$ at 3461~MeV, two more states are obtained, one associated to the bound $D\bar{D}$ state with about 4~MeV binding, and an extra state with large width that could be associated to the $\chi_{c0}(2P)$ state, although it also couples strongly to $D\bar{D}$. This state appears at $3983^{+23}_{-20}$~MeV, which is more than 100~MeV above the claimed $\chi_{c0}(2P)$ state in Ref.~\cite{exp} at 3860~MeV, and more in agreement with the quark model predictions of $3920$~MeV in Ref.~\cite{Godfrey:1985xj}. Although a detailed study is not done in Ref.~\cite{Ortega:2009hj}, the lattice data also contain valuable information to be more precise about the content of the $c\bar{c}$ and molecular components following an analysis as done in Ref.~\cite{Torres:2014vna}, a work which would be further clarifying about the nature of the states.

With the perspective given by the above discussion, we can only encourage our experimental colleagues to measure the reactions studied here with larger statistics and precision, and our theoretical colleagues to pursue work along the line discussed here.

NOTE: After this work was completed, a detailed analysis of the $B^+\to D^+D^- h$ ($h$ an extra hadron) reaction has been conducted, leading to the publication of the papers~\cite{Aaij:2020ypa,Aaij:2020hon}. Analyzing  the invariant mass and angular distributions, two $\chi_{cJ}$ resonances are reported, with the same mass, the $\chi_{c0}(3930)$ and the $\chi_{c2}(3930)$, and widths around 17~MeV and 34~MeV, respectively. Fits with the explicit consideration of $\chi_{c0}(3860)$ of Ref.~\cite{exp} are  conducted and found unfavorable, with the  conclusion ``{\it There is no evidence for the $\chi_{c0}(3860)$ state reported by the Belle Collaboration~\cite{exp}}".

\begin{acknowledgments}
This work is partly supported by the National Natural Science Foundation of China under Grants Nos. 11975083, and 11947413.  It is also supported by the Key Research Projects of Henan Higher Education Institutions under No. 20A140027, Training Plan for Young Key Teachers in Higher Schools in Henan Province (2020GGJS017), the Academic Improvement Project of Zhengzhou University, and the Fundamental Research Cultivation Fund for Young Teachers of Zhengzhou University (JC202041042).
This work is also partly supported by the Spanish Ministerio de Economia y Competitividad
and European FEDER funds under the contract number FIS2011-28853-C02-01, FIS2011-28853-C02-02, FIS2014-57026-REDT, FIS2014-51948-C2-1-P, and FIS2014-51948-C2-2-P.
\end{acknowledgments}

  \end{document}